\documentclass[12pt]{article}
\usepackage[cp1251]{inputenc}
\usepackage{cyr}
\usepackage{epsf}
\usepackage{pazh}
\usepackage[russian]{babel}
\tightenlines

\def\*{$^{*}$}
\def\Б{$^{\mbox{\small Б}}$}
\def\В{$^{\mbox{\small В}}$}
\def\Ч{$^{\mbox{\small Ч}}$}
\def\З{$^{\mbox{\small З}}$}
\def\Д{$^{\mbox{\small Д}}$}
\def\ЕТЗУ{ЬТЗ~У$^{-1}$}
\def\ЕТЗУН{ЬТЗ~УН$^{-2}$~У$^{-1}$}
\def\etal{{et~al.}}
\sloppypar

\begin{document}
\baselineskip 21pt

\title{\bf Extended Emission from Short Gamma-Ray Bursts Detected
with SPI-ACS/INTEGRAL}

\author{\bf \hspace{-1.3cm}\ \
P. Yu. Minaev\affilmark{1,2*}, A. S. Pozanenko\affilmark{1}, V. M.
Loznikov\affilmark{1}}

\affil{
{\it $^1$ Space Research Institute, Russian Academy of Sciences, Profsoyuznaya ul. 84/32, Moscow, 117997 Russia}\\
{\it $^2$ Sternberg Astronomical Institute, Universitetskii pr.
13, Moscow, 119992 Russia}}

\vspace{2mm}

\sloppypar \vspace{2mm} \noindent The short duration ($T_{90}$ < 2
s) gamma-ray bursts (GRBs) detected in the SPI-ACS experiment
onboard the INTEGRAL observatory are investigated. Averaged light
curves have been constructed for various groups of events,
including short GRBs and unidentified short events. Extended
emission has been found in the averaged light curves of both short
GRBs and unidentified short events. It is shown that the fraction
of the short GRBs in the total number of SPI-ACS GRBs can range
from 30 to 45\%, which is considerably larger than has been
thought previously.



\vfill
\noindent\rule{8cm}{1pt}\\
{$^*$ E-mail: $<$minaevp@mail.ru$>$}

\clearpage

\section*{INTRODUCTION}
\noindent

Bimodality in the gamma-ray burst (GRB) duration distribution was
discovered in a series of Konus experiments (Mazets et al. 1981)
and was subsequently confirmed by extensive statistical data from
the BATSE experiment (Kouveliotou et al. 1993), where a robust GRB
duration parameter ($T_{90}$) was proposed (see the \glqq Data
Processing\grqq~Section). The value of $T_{90}$, which separates
the groups of GRBs, can depend on experiment ($T_{90}$ = 2 s is
used in BATSE). Long GRBs have a duration $T_{90}$ > 2 s. Long
GRBs have a softer spectrum than that of short ones; they also
possess a spectral lag - the time profile in softer energy
channels lags behind that in hard ones. Long GRBs are believed to
result from the collapse of massive stars (see, e.g., Paczynski
1998). The nature of short ($T_{90}$ < 2 s) GRBs has not yet been
clarified completely. Theoretical studies showed that short GRBs
could result from the mergers of compact objects (neutron stars or
black holes) in binary systems (Paczynski 1986). This is confirmed
by the absence of observational signatures of a supernova in the
optical afterglow light curves of short GRBs. It is also
speculated that there exist very short ($T_{90}$ < 0.1 s) GRBs
that constitute a separate class of events, along with the
currently identified long and short GRBs, and that the evaporation
of primordial black holes in the Galaxy could be the source of
very short GRBs (Page and Hawking 1976; Cline et al. 2006).

Recently, one of the most distant GRBs has been detected, GRB
090423, with a redshift z $\sim8.2$ and a duration in the
observer's frame of reference $T_{90}$ = 10.3 s, but in the frame
of reference associated with the GRB source, $T_{90}$ = 1.1 s. It
is unknown how this event should be classified, as a long or short
burst, because the significant energy release ($10^{53}$ erg) of
this event is typical of long GRBs. Unfortunately, apart from its
duration $T_{90}$, any additional information about a specific
event (its energy spectrum, the source's redshift, etc.) cannot
always be obtained. Therefore, there exists an uncertainty in
choosing a model of a specific event based only on its duration
$T_{90}$ in the observer's frame of reference. One of the serious
problems of a short event identification with a GRB is related to
the existence of soft gamma repeaters whose light curves are very
similar to those of GRBs (Mazets et al. 2008).

Extended emission (possibly, an afterglow) in the soft gamma-ray
range (>25–50 keV) with a duration of more than 30 s (Table 1) was
found in the averaged light curve of short GRBs in the BATSE
(Lazzati et al. 2001; Connaughton 2002), Konus (Frederiks et al.
2004), and BeppoSAX (Montanari et al. 2005) experiments. Extended
emission with a duration of several tens of seconds was also found
in the light curves of individual short GRBs ($T_{90}$ < 2 s) in
the Swift, BATSE, HETE-2, and Konus experiments (Burenin 2000;
Norris and Bonnell 2006); the extended emission found in these
experiments has a softer spectrum with respect to the main short
episode, which is similar in its spectral properties to short
bursts. A spectral lag is present in the light curves of several
events with extended emission and such GRBs can be classified as
long bursts. But the light curves of most \glqq short\grqq~GRBs
with extended emission exhibit no spectral lag, for example, in
the case of GRB 060614 (Gehrels et al. 2006): the duration
$T_{90}$ in the energy range 15-350 keV is 102 s; no signature of
a supernova was found at the location of GRB 060614, although its
distance is small (z = 0.125). The light curve of this event
consists of a short hard episode with a duration of less than 5 s
and a softer emission with a duration of $\sim100$ s. There is no
spectral lag in the light curve. Is this GRB long or short? Is
extended emission a common property of short GRBs? What is the
physics of the extended emission - is this emission an extension
of the activity of the \glqq central machine\grqq~or this is the
onset of an X-ray afterglow? In this paper, we attempted to answer
these questions by investigating the short GRBs detected in a
harder energy range (>80 keV) with respect to previous studies of
the extended emission with the anticoincidence shield (ACS) of the
SPI spectrometer onboard the INTEGRAL observatory. We also
compiled a catalog of confirmed short GRBs detected with
SPI-ACS/INTEGRAL over the period 2002-2007, which complements and
extends the catalog by Rau et al. (2005) containing data for
2002-2005 in the part of short GRBs.

\section*{DATA SELECTION AND PROCESSING}
\subsection*{SPI-ACS INTEGRAL}
\noindent

SPI-ACS consists of a plastic scintillator PSAC, BGO crystals
(bismuth germanate) of the upper and lower collimator rings and
the lower protective shield; BGO crystals are also located in the
side walls. The BGO crystals are viewed by photomultiplier tubes
(PMTs) and the counts from all PMTs are recorded in a single
channel. SPI-ACS records photons almost from all directions. The
direction coincident with the SPI field of view,
$30^{\circ}\times30^{\circ}$, is least sensitive. SPI-ACS has a
lower sensitivity threshold of $\sim80$~keV - the physical
properties of individual BGO assemblies (detector + PMT +
discriminator) slightly differ and have different lower thresholds
from 60 to 120 keV; the upper threshold is $\sim10$~MeV. The
SPI-ACS time resolution is 50 ms (von Kienlin et al. 2003).
SPI-ACS has a stable background level owing to the high elliptical
orbit of the observatory. Although no spectral information can be
obtained with SPI-ACS, the high upper sensitivity threshold and
the field of view close to $4\pi$ make it a powerful instrument
for investigating the class of hard short GRBs. SPI-ACS is part of
IBAS (INTEGRAL Burst Alert System) (Mereghetti et al. 2003). The
IBAS software algorithm selects events on nine different time
scales (0.05, 0.1, 0.2, 0.4, 0.8, 1, 2, and 5 s), provided that
the event significance with respect to the mean background is 9,
6, 9, 6, 9, 6, 9, and $6\sigma$, respectively. The light curves of
the events selected by this algorithm (containing data from -5 to
100 s relative to the trigger time) are publicly accessible
(http://isdcarc.unige.ch/arc/FTP/ibas/spiacs/).


\subsection*{Catalog of Confirmed SPI-ACS GRBs}
\noindent

We partially used the catalog by Rau et al. (2005), which contains
data on GRBs confirmed by other observatories: event date and time
(UTC), significance, event duration $T_{90}$, fluence and peak
flux on a 0.25-s scale. Apart from the data on confirmed GRBs,
this catalog contains information about unconfirmed events that
are candidates for GRBs. Such events were selected according to
the following algorithm: each event selected by IBAS was checked
for coincidence with the IREM (INTEGRAL Radiation Monitor, Hajdas
et al. 2003) and GOES (Geostationary Operational Environmental
Satellites, http://www.sec.noaa.gov) experiments in order to try
to eliminate the events that result from the interaction of
SPI-ACS with charged particle beams; the soft gamma repeaters
(SGRs) were excluded; in addition, the event significance must
exceed a $12\sigma$ threshold in the original SPI-ACS time
resolution, 0.05 s. Thus, the catalog by Rau et al. (2005)
contains data on 388 events detected from November 27, 2002, to
January 12, 2005, among which only 179 events are confirmed GRBs.
Events with a duration $T_{90}\leq$ 0.05 s (0.05 s is the SPI-ACS
time resolution) make the greatest contribution ($\sim40\%$) to
the group of unconfirmed short events. This group of events is
described in more detail in the \glqq Data Processing\grqq~and
\glqq Discussion\grqq~Sections. We compiled a catalog of confirmed
short GRBs detected with SPI-ACS INTEGRAL from November 27, 2002,
to January 23, 2008, (Table 2), which includes short bursts
confirmed by other observatories. The confirmation search sources
are the electronic GRBlog catalog (http://grad40.as.utexas.edu/
grblog.php), which is a full compilation of GCN circulars
(http://gcn.gsfc.nasa.gov/) and IAU telegrams, and the master
catalog by Hurley (2008)
(http://www.ssl.berkeley.edu/ipn3/masterli.txt).

\subsection*{Data Processing}
\noindent

The catalog (Table 2) contains data on the 83 confirmed GRBs with
a duration $T_{90}$ < 2 s detected with SPI-ACS/INTEGRAL over the
period 2002–2007, which complements and extends the catalog by Rau
et al. (2005) in the part of short bursts. We selected 53 light
curves to construct an averaged light curve, because the full
light curves for the remaining events were inaccessible either due
to their being not public or due to the stopping of telemetry
because of the observatory’s slew. These GRBs constitute the first
group of investigated events. The second group consists of short
bursts unconfirmed by other space observatories, candidates for
GRBs from the catalog by Rau et al. (2005) with a duration
$T_{90}$ < 2 s (105 events). The third group is a subset of the
second group - this includes short unconfirmed candidates for GRBs
from the catalog by Rau et al. (2005) with a duration
$T_{90}\leq$0.05 s (43 events).

As was shown by Rau et al. (2005), there exists a class of short
events with a duration $T_{90}\leq$0.05 s detected with SPI-ACS
that has the following property: saturation, the absence of any
signal for several (up to several tens) seconds, is observed in
one or more neighboring SPI germanium detectors simultaneously
with the short burst in SPI-ACS. The light curve for one of such
events is shown in Fig. 1. It is suggested (Rau et al. 2005) that
this class of events is related to the interaction of the SPI-ACS
and SPI detectors with charged particles. The fourth, test group
consists of such events (the first, second, and third groups
contain no such events).

All light curves were aligned relative to the main peak using a
procedure similar to that proposed by Mitrofanov et al. (1996) (as
an example, Fig. 2 shows the light curve of GRB 060221) and were
investigated in the time interval [-150; 200] s. The processing
procedure consists of the following steps:

(1) Background model approximation. The background variations on
various scales were investigated: we grouped the bins with a time
resolution of 50 s, as shown in Fig. 3 for revolution 405 of the
observatory, and calculated the maximum background changes on a
time scale of 350 s corresponding to the interval in which the
light curves of all the events being analyzed were investigated.
It follows from Fig. 3 that the behavior of the background is
monotonic on the 350-s time scale. Therefore, we used a linear
background model in the intervals [-150;-50] and [100; 200] s
relative to the main peak. We also established that the background
in the 350-s time interval changed by no more than 0.3\%. In an
absolute majority of cases, the model describes well the
background behavior. However, in several cases, the quality of the
linear model fit was unsatisfactory and these events were excluded
from further consideration. Our analysis of the background
variance confirmed the deviation of the statistical signal
distribution from a Poisson law — the $1\sigma$ significance of
the signal above the background B is determined from the formula
$1.57\times B^{1/2}$, which independently confirms the papers by
von Kienlin et al. (2003) and Ryde et al. (2003). All the
subsequent calculations of the standard deviations take this fact
into account.

(2) Background model subtraction from the light curve.

(3) Calculation of the event duration $T_{90}$ (the time in which
90\% of the burst energy is emitted). The algorithm for
calculating $T_{90}$ is discussed in Koshut et al. (1996). Figure
4 presents the scheme for calculating $T_{90}$ for GRB 030325. An
integral light curve of the event is constructed. Subsequently,
the numbers of counts corresponding to 5\% and 95\% of the total
number of counts (indicated in Fig. 4 by the horizontal solid
lines) are determined. Next, the times $T_{5}$ and $T_{95}$
(vertical solid lines) corresponding to these fluxes are found and
their difference, which is the duration $T_{90}$ = $2.00\pm0.15$
s, is calculated. The values of $T_{90}$ we calculated were
compared with those from Rau et al. (2005) and with the values
published for the same events but determined in different
experiments: the durations in the RHESSI experiment were taken
from Ripa et al. (2009), in the Suzaku experiment from the catalog
at http://www.astro.isas.jaxa.jp/suzaku/HXDWAM/ WAM-GRB/, in case
of the Swift experiment from Sakamoto et al. (2008), and, in the
remaining cases, from GCN circulars. These values are also given
in Table 2. No significant differences in durations were found.

(4) Grouping of bins — increasing the bin duration from 0.05 to 5
s. Figure 5 shows the result of processing the light curve for GRB
060221.

(5) Averaging. The light curves of all GRBs aligned relative to
the main peak and processed according to the procedure described
above were averaged (the light-curve points corresponding to the
same time in different light curves were averaged). In the
individual light curves of GRB 060221 (Fig. 5) and GRB 031214
(Fig. 6) after the processing procedure described above, we found
statistically significant extended emission with a duration of
$\sim50$ and $\sim20$ s ($T_{90}$ for these events are 0.2 and 0.3
s, respectively), with the extended emission from GRB 031214
having been also found in the Konus experiment (Oleynik et al.
2008). These two GRBs were excluded from the subsequent analysis.

As has already been noted in the Introduction, it is often
impossible to determine whether a GRB belongs to the class of
short or long bursts based only on its light curve. Studies of the
event spectrum, analysis of the lag in various energy channels,
and a recently proposed method for burst separation in Amati’s
$E_{p} - E_{iso}$ diagram (Amati 2010) are needed for the most
comprehensive analysis. There is no clear boundary in the duration
distribution either and the boundary $T_{90}$ = 2 s was chosen by
analogy with BATSE. Therefore, events belonging to the class of
long bursts can also be present in the sample of bursts. The
number of such events can be estimated from the lognormal
distributions that describe a bimodal duration distribution. For
BATSE, such estimation for the 2-s boundary gives 2\%. A more
accurate determination of the boundary is discussed, for example,
in Donaghy et al. (2006).

In the averaged light curves for the first and second groups of
short events (51 and 105 events, respectively), we also found
extended emission with a duration of $\sim25$ s (Minaev et al.
2009). Figure 7 presents the light curves of confirmed and
unconfirmed short events (the first and second groups of GRBs).
Since extended emission was found in the averaged light curve of
short GRBs in various experiments (SPI-ACS/INTEGRAL, BATSE, Konus,
BeppoSAX), it can be assumed that this is an actually existing
phenomenon. From the available data for short GRBs, we can draw
only statistical conclusions based on the averaging of various
samples (confirmed and unconfirmed GRBs). We cannot assert that
extended emission is a property of each short GRB. Nevertheless,
the extended emission found in the averaged light curve of
unconfirmed short events suggests that some of the
\textit{unconfirmed} short events belong to the class of real
GRBs. For an independent test, we selected short events associated
with triggers from charged particles (the fourth group of GRBs).
There is no extended emission in the averaged light curve of 33
such events (Fig. 8). Consequently, we have no reason to reject
the assumption that some of the unconfirmed short events are real
GRBs.

\section*{DISCUSSION AND CONCLUSIONS}
\noindent

In the averaged light curve for the first group of short events
(confirmed GRBs), we found extended emission with a duration of 25
s and an intensity of $(46\pm15)$ counts/s (Fig. 7). Thus,
extended emission can be assumed to be a common property of all
short GRBs: it was found in the averaged light curve of the short
GRBs detected in various experiments (see Table 1). The extended
emission found in the averaged light curve of unconfirmed short
events (the second group) suggests that some of the events from
this group also belong to the class of real GRBs and not to
triggers from charged particles and that the total fraction of the
short GRBs is considerably higher than has been thought
previously. If the intensity of the extended emission is assumed
to be the same for all short GRBs, then the number of real short
GRBs in the group of unconfirmed events can be estimated from the
intensity of the extended emission in the averaged light curves.
Thus, the fraction of the real GRBs in the group of unconfirmed
events is $(84\pm35)$\%.

The fraction of the short GRBs detected in BATSE is $\sim25\%$ of
all the detected GRBs (Kouveliotou et al. 1993), their fraction in
the APEX experiment is $\sim38\%$ (Kozyrev et al. 2004), and the
fraction of the short GRBs detected with SPI-ACS/INTEGRAL and
confirmed by other space observatories is 16\%. If all unconfirmed
short events (the second group) are assigned to the class of real
GRBs, then the total fraction of the short GRBs detected with
SPI-ACS/INTEGRAL will be $\sim45\%$, which is the upper limit for
the number of short GRBs in the SPI-ACS experiment. The lower
limit is 30\% under the assumption that the fraction of the real
GRBs in the group of unconfirmed events is 84\%.

Since the lower energy threshold for BATSE triggers (for most part
of the mission $\sim50$ keV) is lower than that for SPI-ACS, the
fraction of the harder short events detected with SPI-ACS is more
than 25\%. This is an additional argument for the hypothesis that
some of the unconfirmed short events are real cosmic GRBs. The
smallest trigger windows (the time intervals in which the signal
is compared with the background value) for BATSE and SPI-ACS are
approximately the same, 64 and 50 ms, respectively. It is also
interesting to estimate the sensitivity of SPI-ACS to short events
and to compare it with the corresponding sensitivity of BATSE.
Vigano and Mereghetti (2009) discuss the conversion of SPI-ACS
counts to energy units: for a normal angle between the source and
the X axis of the INTEGRAL observatory, 1 SPI-ACS count
corresponds to $10^{-10}$ erg cm$^{-2}$. Taking into account the
mean background value in SPI-ACS and the trigger algorithm, we
will find that the sensitivity of SPI-ACS to short events is
$\sim6\times10^{-8}$ erg cm$^{-2}$ and $\sim 1.7\times10^{-7}$ erg
cm$^{-2}$ on time scales of 50 ms and 1 s, respectively. The
sensitivity of BATSE on a time scale of 1 s is $\sim 10^{-7}$ erg
cm$^{-2}$ (Fishman et al. 1994).

To investigate the extended emission, we fitted the averaged light
curve for the combined first and second groups of events by
functions $y=at^{b}$ and $y=ce^{dt}$ in the segment [0; 40] s
relative to the main peak with a time resolution of time profile
of 5 s. Figure 9 presents the signal significance for the combined
groups. To estimate the likelihood of these models, we used the
$\chi^{2}$ test. The values obtained (with six degrees of freedom)
are 11.8 and 16.2 for the first and second models, respectively.
Consequently, the significance that the extended emission is
described by these models is low, no more than 7\% and 1\%,
respectively. The power-law model of the extended emission cannot
be rejected, while the exponential model describes the behavior of
the extended emission considerably more poorly, which probably
rules out the FRED model of the extended emission. On the other
hand, since the extended emission was found in a hard energy range
($\geq80$ keV), the extended emission found can be assumed to be
an extension of the activity of the GRB central machine rather
than the onset of an X-ray afterglow.

In the averaged light curve for the third group of events
(unconfirmed short GRBs with a duration $T_{90}\leq$0.05 s), we
also found extended emission with a duration of 125 s and a total
intensity of $(213\pm35)$ counts/s (Fig. 6). This suggests that
the nature of \glqq very short\grqq~GRBs is probably the same as
that of \glqq ordinary\grqq~short GRBs with a duration
$T_{90}\leq$2 s (the group of very short GRBs is studied in more
detail in Minaev et al. (2010)). However, it is speculated (Cline
et al. 2006) that very short GRBs result from the evaporation of
hypothetical primordial black holes. The light curve corresponding
to the evaporation of a primordial black hole consists of a very
short episode whose duration is fractions of a second (Cline et
al. 2006; Petkov et al. 2008) and no extended emission must be
observed from such events. Therefore, the very short GRBs detected
with SPI-ACS/INTEGRAL probably should not be assigned to the class
of events related to the evaporations of primordial black holes.
There is no extended emission in the averaged light curve for the
test, fourth group of events (triggers from charged particles)
(Fig. 8).

For the groups being investigated, we constructed a cumulative
$log N-log C_{max}$ distribution (Fig. 10), where $N$ is the
number of bursts with a count rate at the light-curve maximum
exceeding $C_{max}$. The deviation from uniform distribution of
sources in 3D Euclidean space of the curve for the first group of
events (confirmed short GRBs from our catalog) at low values of
$C_{max}$ can be attributed to the selection effect: the
low-intensity events are missed. The curve for the sample of
unconfirmed GRBs is closest to the uniform distribution of sources
in Euclidean space (the \glqq -3/2 law\grqq). Therefore, if the
unconfirmed short GRBs are assumed to have the same nature as that
of the confirmed ones, then the overall $log N-log C_{max}$
distribution will satisfy the -3/2 law. This corroborates the
present views of the spatial distribution of short GRBs as objects
of the close Universe (z < 1, Gehrels 2008).

\section*{ACKNOWLEDGMENTS}
We wish to thank S.A. Grebenev, S.V. Mol’kov, and I.V. Chelovekov
for a helpful discussion of the work and their remarks. This work
was supported by the \glqq Origin, Structure, and Evolution of
Objects in the Universe\grqq~Program of the Russian Academy of
Sciences.

 \noindent

 \pagebreak

\clearpage
\begin{table}[t]

\vspace{6mm} \centering {{\bf Table 1.} Extended emission in the
averaged light curve of short GRBs}\label{meansp}

\vspace{5mm}\begin{tabular}{l|c|c|c} \hline\hline
{Experiment}&Energy &Number of &Emission  \\
             &range, keV&investigated         &duration, s\\
             &          &GRBs& \\ \hline
BATSE &25-110              &76&    100   $^{1}$    \\
BATSE &50-300   &100             &  100 $^{2}$\\
Konus   &10-750&125&     100 $^{3}$\\
BeppoSAX   &40-700&   93          &   30 $^{4}$\\
INTEGRAL   &$>80$&     53        &25 $^{5}$\\
INTEGRAL   &$>80$&     43        &125 $^{6}$\\ \hline
\multicolumn{4}{l}{}\\ [-3mm]
\multicolumn{4}{l}{$^{1}$ - Lazzati et al. (2001).}\\
\multicolumn{4}{l}{$^{2}$ - Connaughton (2002).}\\
\multicolumn{4}{l}{$^{3}$ - Frederiks et al. (2004).}\\
\multicolumn{4}{l}{$^{4}$ - Montanari et al. (2005).}\\
\multicolumn{4}{l}{$^{5}$ - This paper, the first group of events.}\\
\multicolumn{4}{l}{$^{6}$ - This paper, the third group of events.}\\
\end{tabular}
\end{table}

\begin{table}[t]

\vspace{6mm} \centering {{\bf Table 2.}  Catalog of confirmed
short GRBs detected with SPI-ACS/INTEGRAL. }\label{meansp}

\vspace{5mm}\begin{tabular}{l|c|c|c|c|c|c} \hline\hline
{GRB}    &Trigger          &$T_{90}$,                  &$C_{max}$,         &Confirmation $^{1}$          &$T_{90}$,    &Comm. \\
         &time, UT    &   s                       &$10^{3}$cnts      &                        & s $^{(2)}$    &               \\ \hline
030101   &   20:43:32    &   0.7     $\pm$   0.1     &   1.9 $\pm$   0.1 &   u, h, k              & 1.0[1]     &   $^{+}$    \\
030105   &   14:34:12    &   1.2     $\pm$   0.15**    &   3.7 $\pm$   0.1 &   m, k, r            & 1.23[2]       &   $^{+}$    \\
030109   &   9:37:37     &   0.35    $\pm$   0.1     &   0.6 $\pm$   0.1 &   u                    &      &    $^{+}$   \\
030110   &   9:39:28     &   0.1     $\pm$  0.05     &   1.3 $\pm$   0.1 &   k                     &     &  $^{+}$ sl.$^{3}$\\
         &               &                           &                   &                         &     & (130s)$^{4}$    \\
030117   &   17:36:14    &   0.15    $\pm$   0.05    &   3.7 $\pm$   0.1 &   k                     &     &  $^{+}$     \\
030217   &   23:31:42    &   0.35    $\pm$   0.05    &   4.1 $\pm$   0.1 &   m                     &     &   $^{+}$    \\
030325   &   14:15:11    &   2       $\pm$   0.15    &   2   $\pm$   0.1 &   u, k, m            &2.0[3] &    $^{+}$   \\
030523   &   14:10:52    &   0.15    $\pm$   0.05    &   1   $\pm$   0.1 &   k, r        &0.12[2]       &   $^{+}$    \\
030607   &   2:19:21     &   0.1     $\pm$   0.05    &        -           &   k             &             &   sl.    \\
030629   &   3:26:39     &   0.15    $\pm$   0.05    &   3.1 $\pm$   0.1 &   k             &             & $^{+}$  sl.  \\
         &               &                           &                   &                 &             & (-95s)    \\
030711   &   0:04:01     &   0.25    $\pm$   0.05    &       -            &   m              &            &   sl.   \\
030717   &   20:49:24    &   0.05    $\pm$   0.05    &   4.1 $\pm$   0.1 &   u             &0.06[4]     & bad bgd. $^{5}$    \\
030916   &   21:59:18    &   0.65    $\pm$   0.1     &   1   $\pm$   0.1 &   k          &                &   $^{+}$    \\
030926   &   16:52:27    &   0.2     $\pm$   0.05    &   0.6 $\pm$   0.1 &   r, h, m, k     &0.28[2]         &    $^{+}$   \\
030929   &   14:27:14    &   0.45    $\pm$   0.05    &      -             &   k              &            &   sl.    \\
031026   &   1:26:29     &   0.25    $\pm$   0.05    &   2.3 $\pm$   0.1 &   m, k            &           &   $^{+}$    \\
031208   &   1:18:28     &   2.0     $\pm$   0.5     &        -           &   k               &           &   sl.    \\
031210   &   11:51:06    &   0.7     $\pm$   0.1     &   0.4 $\pm$   0.1 &   k               &           &    $^{+}$   \\
031214   &   10:10:50    &   0.3     $\pm$   0.05    &   59.9$\pm$   0.4 &   k, m           &0.3[5]    &    ext. em. $^{6}$   \\
040202   &   13:29:52    &   0.4     $\pm$   0.05    &   2   $\pm$   0.1 &   u              &            &    $^{+}$   \\
040312   &   0:02:35     &   0.3     $\pm$   0.1     &   2.9 $\pm$   0.1 &   k, r           &0.16[2]           &    $^{+}$   \\
040322   &   7:29:02     &   0.1     $\pm$   0.05    &        -           &   k, m           &0.19[6]   &   sl.   \\
040324   &   10:21:10    &   0.2     $\pm$   0.05    &   11.7$\pm$   0.2 &   k, r           &0.26[2]          & $^{+}$      \\
040329   &   11:10:49    &   2       $\pm$   0.05    &   14.3$\pm$   0.2 &   k, r           &2.07[2]           &   $^{+}$    \\
040417   &   8:05:09     &   1.25    $\pm$   0.1     &   0.4 $\pm$   0.1 &   k              &                      &    $^{+}$   \\
040802   &   18:02:20    &   1.1     $\pm$   0.3     &   0.5 $\pm$   0.1 &   h, m, k        &                    &   $^{+}$    \\
040822   &   21:21:53    &   0.55    $\pm$   0.1     &   0.9 $\pm$   0.1 &   k, r           &1.38[2]          &   $^{+}$    \\
041013   &   22:56:26    &   0.35    $\pm$   0.05    &   2.2 $\pm$   0.1 &   m, k, r        &0.36[2]            &   $^{+}$    \\
041116   &   14:42:41    &   1.1     $\pm$   0.15    &   1   $\pm$   0.1 &   h, k           &0.5[7]      &   $^{+}$    \\
041213   &   6:59:36     &   0.1     $\pm$   0.05    &   2.9 $\pm$   0.1 &   k, r           &0.14[2]               &  $^{+}$ sl.    \\
         &               &                           &                   &                  &            & (100s)    \\
050111   &   6:52:26     &   0.1     $\pm$   0.1     &   0.9 $\pm$   0.1 &   k, h           &            &   $^{+}$    \\
050112   &   11:10:23    &   0.45    $\pm$   0.05    &   17.9$\pm$   0.2 &   k, h           &0.52[8]    &    $^{+}$   \\
050212   &   21:24:12    &   0.2     $\pm$   0.05    &   6.9 $\pm$   0.2 &   k              &           &   $^{+}$    \\
050216   &   7:26:34     &   0.3     $\pm$   0.05    &   2.9 $\pm$   0.1 &   r, k           &0.5[2]    &   $^{+}$    \\

\end{tabular}
\end{table}
\clearpage

\begin{table}[t]

\vspace{6mm} \centering {{\bf Table 2.}  (Contd.) }\label{meansp}

\vspace{5mm}\begin{tabular}{l|c|c|c|c|c|c} \hline\hline
{GRB}    &Trigger         &$T_{90}$,                  &$C_{max}$,         &Confirmation                 &$T_{90}$,     &Comm. \\
         &time, UT   &   s                       & $10^{3}$cnts.     &                              &      s &       \\ \hline

050328   &   3:25:14     &   -                   &       -            &   k, r               &0.45[2]               &   sl.    \\
050409   &   1:18:35     &   1.15    $\pm$   0.1     &   10.6$\pm$   0.2 &   m, k, r        &1.26[2]            &  $^{+}$     \\
050502   &   19:56:55    &   1.05    $\pm$   0.05    &   0.4 $\pm$   0.1 &   r              &1.6[2]                   &   $^{+}$ sl.    \\
         &               &                           &                   &                  &            & (185s)    \\
050513   &   4:39:59     &   0.8     $\pm$   0.1     &   1.3 $\pm$   0.1 &   k, m, mes                  &       &    $^{+}$    \\
050724   &   12:34:09    &   0.75    $\pm$   0.15    &   0.9 $\pm$   0.1 &   k, s                       &0.25[9]     &    $^{+}$    \\
050805   &   13:29:47    &   0.45    $\pm$   0.05    &   2.8 $\pm$   0.1 &   r, k, m                    &1.05[2]       &    $^{+}$    \\
050809   &   20:15:24    &   1.35    $\pm$   0.15    &   1.5 $\pm$   0.1 &   r, k, m                    &2.4[2] &    $^{+}$    \\
050821   &   10:55:41    &   0.2     $\pm$   0.05    &   1.9 $\pm$   0.1 &   k, sz                   &2.0[10]       &     $^{+}$   \\
051016   &   5:19:37     &   1.05    $\pm$   0.05*    &       -            &   k                       &           &   sl.   \\
051107   &   2:30:41     &   -                   &     -              &   k, s $^{7}$, sz                &1.75[11],5[10]    &   sl.    \\
051221   &   1:51:15     &   0.25    $\pm$   0.05    &   9.1 $\pm$   0.2 &   m, r, s, sz            &0.28[2],1.4[12], &   $^{+}$     \\
&&&&&0.5[10]&\\
060103   &   8:42:47     &   0.9     $\pm$   0.1     &   0.8 $\pm$   0.1 &   m, k                     &          &    $^{+}$    \\
060126   &   9:30:04     &   0.15    $\pm$   0.05    &   1.9 $\pm$   0.1 &   h, r                     &          &    $^{+}$    \\
060130   &   6:10:52     &   -                   &         -          &   k, r, sz           &0.41[10]                   &  bad bgd.     \\
060221   &   21:14:58    &   0.2     $\pm$   0.15    &   0.8 $\pm$   0.1 &   sz                     &0.75[10]                         &   ext. em.    \\
060303   &   22:42:47    &   0.35    $\pm$   0.05    &   9.5 $\pm$   0.2 &   k, s $^{7}$, sz, r     &0.38[10], 0.5[2]         &   $^{+}$     \\
060306   &   15:22:39    &   0.85    $\pm$   0.05    &   64.1$\pm$   0.4 &   k, r, s $^{7}$         &0.92[2]                 &   $^{+}$ sl.   \\
         &               &                           &                   &                          &           & (-90с)    \\
060312   &   6:17:20     &   0.6     $\pm$   0.3     &   0.6 $\pm$   0.1 &   r, s $^{7}$, sz        &3[10], 0.24[2]             &    $^{+}$    \\
060313   &   0:12:06     &   0.7     $\pm$   0.05    &   6.4 $\pm$   0.2 &   m, k, s                &0.74[12], 0.8[13]    &    $^{+}$    \\
060425   &   16:57:37    &   -                   &     -              &   s $^{7}$, sz, r    &0.14[2],0.19[10]       &   sl.    \\
060427   &   23:51:55    &   -                   &        -           &   m, k, s $^{7}$      &0.2[14]          &   sl.   \\
060429   &   12:19:51    &   0.1     $\pm$   0.05    &   11.9$\pm$   0.2 &   m, k, r, sz            &0.2[2], 0.13[10]         &      $^{+}$  \\
060601   &   7:55:39     &   -                    &         -          &   sz                 &0.5[10]                         &   sl.    \\
060610   &   11:22:24    &   0.65    $\pm$   0.05    &   2   $\pm$   0.1 &   m, k, r, sz            &0.69[10], 0.6[2]         &   $^{+}$     \\
060823   &   8:05:33     &   1.05    $\pm$   0.05    &   0.7 $\pm$   0.1 &   sz, r                  &1.0[2], 2.0[10]         &  $^{+}$      \\
060912   &   18:31:01    &   -                    &         -          &   sz                 &0.5[10]                          &   n/d$^{8}$ \\
060916   &   14:33:34    &   -                   &          -         &   sz                 &0.13[10]                         &   n/d \\
061001   &   21:14:28    &   0.85 $\pm$ 0.65*        &         -          &   s $^{7}$               &                 &   n/d \\
061003   &   12:14:20    &   0.85 $\pm$ 0.05*        &          -        &   k, s $^{7}$            &                  &   n/d \\
061006   &   16:45:28    &   0.45    $\pm$   0.05    &   2.2 $\pm$   0.1 &   k, r, s, sz            &0.44[10],0.4[2],         &   $^{+}$     \\
&&&&&129.9 [12]$^{9}$&\\
061006   &   8:43:36     &   1.65    $\pm$   0.05    &   2   $\pm$   0.1 &   m, k, r, s $^{7}$, sz &1.6[10], 1.65[2]     &   $^{+}$     \\
061014   &   6:17:02     &   1       $\pm$   0.05    &   2.1 $\pm$   0.1 &   r, sz                &1.2[10], 0.2[2]                &     $^{+}$   \\
061021   &   18:29:24    &   1.1     $\pm$   0.1     &   0.5 $\pm$   0.1 &   k                     &             &    $^{+}$    \\

\end{tabular}
\end{table}
\clearpage

\begin{table}[t]

\vspace{6mm} \centering {{\bf Table 2.}  (Contd.) }\label{meansp}

\vspace{5mm}\begin{tabular}{l|c|c|c|c|c|c} \hline\hline
{GRB}    &Trigger          &$T_{90}$,                 &$C_{max}$,         &Confirmation & $T_{90}$,                &Comm. \\
         &time, UT    &   s                      &$10^{3}$cnts.      &               & s                &       \\ \hline
070113   &   11:56:23    &   -                  &          -         &   k, r, sz           &0.23[10], 0.27[2]            &   n/d \\
070129   &   22:09:25    &   -                   &         -          &   k, s $^{7}$, sz    &0.22[10]                   &   n/d \\
070203   &   23:06:44    &   -                   &          -         &   sz                 &0.69[10]                        &   n/d \\
070321   &   18:52:15    &   0.4     $\pm$   0.05    &   0.7 $\pm$   0.1 &   sz, s $^{7}$, k, m     &0.34[10]               &     $^{+}$   \\
070413   &   20:37:55    &   -                   &         -          &   sz                 &0.19[10]                         &   n/d \\
070516   &   20:41:24    &  -                 &          -         &   k, r, s $^{7}$, mes, sz &1.0[10], 0.35[2]      &   n/d \\
070713   &   13:08:37    &   0.55 $\pm$ 0.2*                   &        -           &   k, s $^{7}$  &                     &   n/d \\
070721   &   14:24:09    &   0.9 $\pm$ 0.25*                   &         -          &   m, k          &              &   n/d \\
070915   &   8:34:48     &   0.65 $\pm$ 0.05*                     &       -            &   k, s $^{7}$, mes &                  &   n/d \\
070921   &   9:47:54     &   1.25 $\pm$ 0.50*                   &         -          &   sz          &2.7[10]                   &  n/d \\
070927   &   22:25:20    &   0.8     $\pm$   0.3    &   0.5 $\pm$   0.1 &   m            &               &    $^{+}$    \\
070927   &   16:27:55    &   0.25    $\pm$   0.15   &   1.4 $\pm$   0.1 &   r, s $^{7}$, mes, sz &              &   $^{+}$     \\
071112   &   18:23:31    &   -                     &          -         &   s, sz            &1.0[10],1.0[15]   &   n/d \\
071227   &   20:13:47    &   -                   &          -        &   k, s, sz         &1.8[16]   &   n/d \\
080121   &   21:29:55    &   -                  &           -        &   s                &0.7[17]          &   n/d \\
080123   &   4:21:57     &   -                   &           -        &   k, s, sz        &0.4[18]   &   n/d \\

\multicolumn{7}{l}{}\\ [-3mm]
\multicolumn{7}{l}{Notes.}\\
\multicolumn{7}{l}{* The values were obtained from public IBAS data, i.e., in the time interval [-5; 100] s }\\
\multicolumn{7}{l}{relative to the trigger time.}\\
\multicolumn{7}{l}{** Ryde et al. (2003) gives a duration of 13 s for this burst. In other sources, the burst}\\
\multicolumn{7}{l}{duration is 0.9 s (Rau et al. 2005) and 1.3 s
(Ripa
et al. 2009).}\\
\multicolumn{7}{l}{In the standard analysis performed in our paper, we found no statistically significant}\\
\multicolumn{7}{l}{extended emission in the SPI-ACS energy
range. A joint analysis of the duration}\\
\multicolumn{7}{l}{and hardness of this burst in the RHESSI experiment (Ripa et al. 2009) allows it to be}\\
\multicolumn{7}{l}{classified
as a short/hard burst with the possible presence of extended emission.}\\
\multicolumn{7}{l}{$^{+}$ Used in the averaged light curve.}\\
\multicolumn{7}{l}{$^{1}$ The observatories that also detected this event: u - Ulysses, h - HETE-2, k - Konus,}\\
\multicolumn{7}{l}{s - Swift, mes - Messenger, sz - Suzaku, r - RHESSI, m - Mars-observer.}\\
\multicolumn{7}{l}{$^{2}$ The durations determined in other experiments, the number of the corresponding reference}\\
\multicolumn{7}{l}{at the end of Table 2 is given in square brackets.}\\
\multicolumn{7}{l}{$^{3}$ The data for the corresponding event are inaccessible due to the observatory’s slew}\\
\multicolumn{7}{l}{(hereafter, \glqq sl.\grqq).}\\
\multicolumn{7}{l}{$^{4}$ The final (or initial) time in the available light curve (relative to $T_{0}$) used}\\
\multicolumn{7}{l}{for averaging is specified.}\\
\multicolumn{7}{l}{$^{5}$ The light curve was excluded from the averaged light curve due to an unsatisfactory}\\
\multicolumn{7}{l}{background fitting quality (hereafter, \glqq bad bgd.\grqq).}\\
\multicolumn{7}{l}{$^{6}$ A short GRB with detected extended emission (hereafter, \glqq ext. em.\grqq).}\\
\multicolumn{7}{l}{$^{7}$ This event was observed outside the BAT/Swift field of view.}\\

\end{tabular}
\end{table}

\clearpage

\begin{table}[t]

\vspace{6mm} \centering {{\bf Table 2.} (Contd.) }\label{meansp}

\vspace{5mm}\begin{tabular}{l|c|c|c|c|c}
\multicolumn{6}{l}{$^{8}$ The original data for the corresponding event are inaccessible (hereafter, \glqq n/d\grqq).}\\
\multicolumn{6}{l}{$^{9}$ The duration of this event is 120 s in the range 15–100 keV and 0.58 s at >100 keV (Lin}\\
\multicolumn{6}{l}{et al. 2008), the burst consists of a short hard peak and a period of activity in the soft}\\
\multicolumn{6}{l}{energy range. This burst was initially classified as a short/hard one (Krimm et al. 2006).}\\
\multicolumn{6}{l}{In our standard analysis, we found no extended emission from GRB 061006 in the SPI-ACS}\\
\multicolumn{6}{l}{energy range and also classify it as a short burst.}\\

\multicolumn{6}{l}{[1] - K. Hurley, E. Mazets, S. Golenetskii
\etal, GRB Coordinates Network {\bf 1783}, 1 (2003).}\\

\multicolumn{6}{l}{[2] - J. Ripa, A. Meszaros, C.
Wigger \etal, Astron. Astrophys. {\bf 498}, 399 (2009).}\\

\multicolumn{6}{l}{[3] - K. Hurley, T. Cline, I. Mitrofanov \etal,
GRB Coordinates Network {\bf 1962}, 1 (2003).}\\
\multicolumn{6}{l}{[4] - K. Hurley, T. Cline, A. von Kienlin
\etal, GRB Coordinates Network {\bf 2307}, 1 (2003).}\\

\multicolumn{6}{l}{[5] - S. Golenetskii, E. Mazets, V. Pal'shin
\etal, GRB Coordinates Network {\bf 2487}, 1
(2003).}\\

\multicolumn{6}{l}{[6] - S. Golenetskii, E. Mazets, V. Pal'shin
\etal, GRB Coordinates Network {\bf 2552}, 1
(2004).}\\

\multicolumn{6}{l}{[7] - S. Golenetskii, R. Aptekar, E. Mazets
\etal, GRB Coordinates Network {\bf 2835}, 1
(2004).}\\

\multicolumn{6}{l}{[8] - S. Golenetskii, R. Aptekar, E. Mazets
\etal, GRB Coordinates Network {\bf 2949}, 1
(2004).}\\

\multicolumn{6}{l}{[9] - S. Covino, L. A. Antonelli, P. Romano
\etal, GRB Coordinates Network {\bf 3665}, 1
(2005).}\\

\multicolumn{6}{l}{[10]
http://www.astro.isas.jaxa.jp/suzaku/HXD-WAM/WAM-GRB/}\\

\multicolumn{6}{l}{[11] - S. Golenetskii, R. Aptekar, E. Mazets
\etal, GRB Coordinates Network {\bf 4234}, 1
(2005).}\\

\multicolumn{6}{l}{[12] - T. Sakamoto, S. D.
Barthelmy, L. Barbier \etal, \apj\ SS {\bf 175}, 179 (2008).}\\

\multicolumn{6}{l}{[13] - S. Golenetskii, R. Aptekar, E. Mazets
\etal, GRB Coordinates Network {\bf 4881}, 1
(2006).}\\

\multicolumn{6}{l}{[14] - S. Golenetskii, R. Aptekar, E. Mazets
\etal, GRB Coordinates Network {\bf 5030}, 1
(2006).}\\

\multicolumn{6}{l}{[15] - M. Perri, S. D. Barthelmy, J. R.
Cummings \etal, GRB Coordinates Network {\bf 7058}, 1}\\
\multicolumn{6}{l}{(2007).}\\

\multicolumn{6}{l}{[16] - G. Sato, L. Barbier, S. D. Barthelmy
\etal, GRB Coordinates Network {\bf 7148}, 1
(2007).}\\

\multicolumn{6}{l}{[17] - J. R. Cummings, and D. M. Palmer, GRB
Coordinates Network {\bf 7209}, 1
(2008).}\\

\multicolumn{6}{l}{[18] - T. Uehara, M. Ohno, T. Takahashi \etal,
GRB Coordinates Network {\bf 7223}, 1
(2008).}\\

\end{tabular}
\end{table}

\clearpage

\begin{figure}[h]
\epsfxsize=19cm \hspace{-2cm}\epsffile{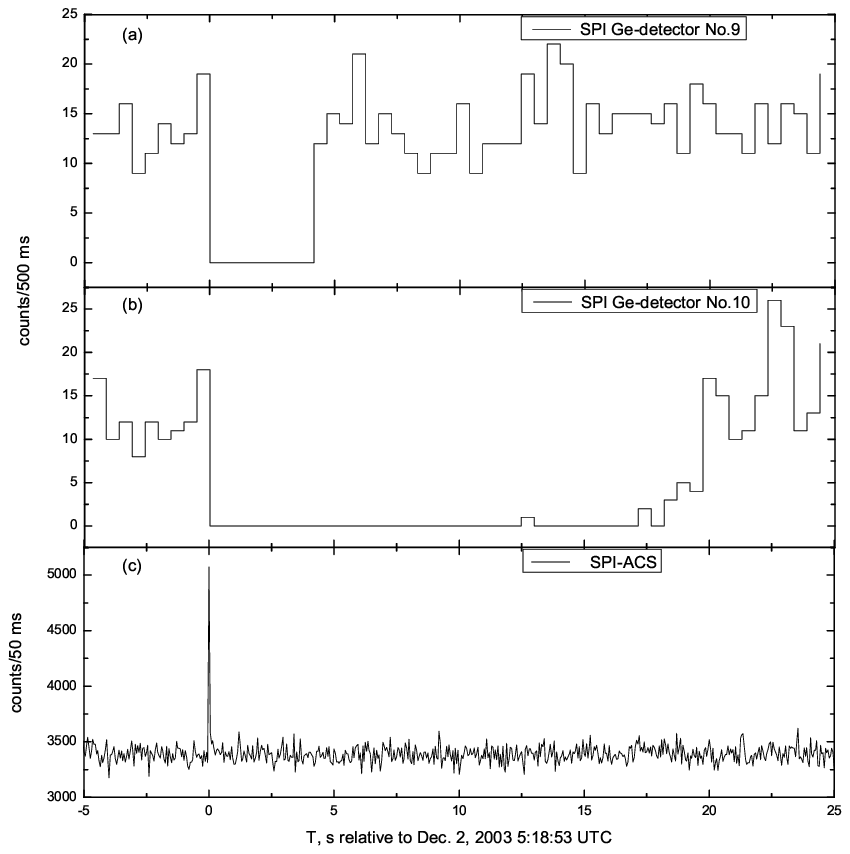} Fig. 1. Example
of a very short SPI-ACS event (c) with simultaneous saturation of
two SPI germanium detectors (a, b). The SPI-ACS light curve is
presented with the original time resolution of 50 ms; the light
curves of the SPI germanium detectors no. 9 and no. 10 are
presented with a time resolution of 500 ms.

\end{figure}

\begin{figure}[h]
\epsfxsize=19cm \hspace{-2cm}\epsffile{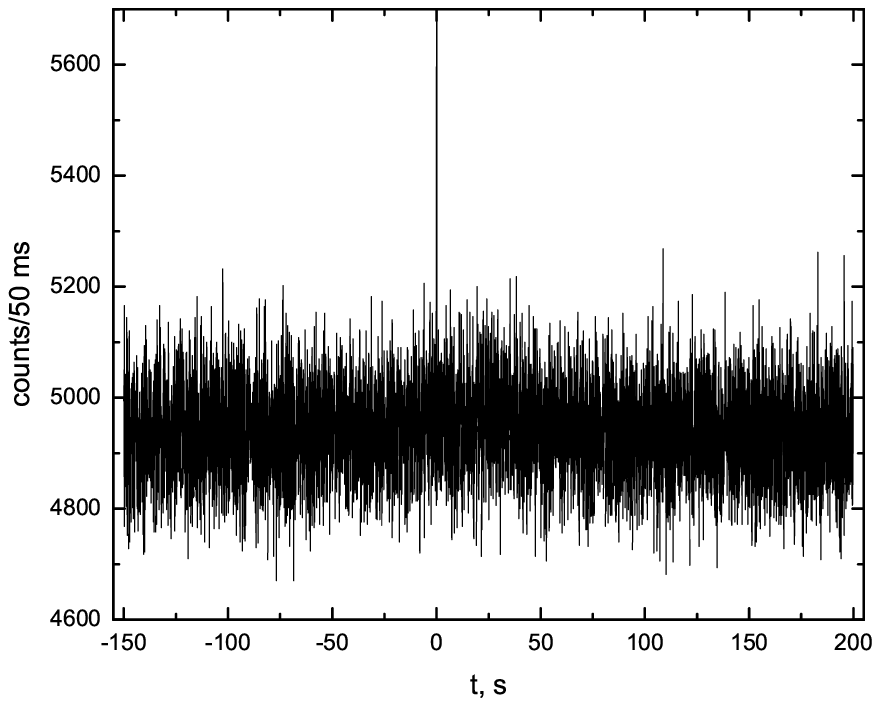} Fig. 2.
Light curve of GRB 060221 with the original SPI-ACS time
resolution of 50 ms. Time relative to the trigger $T_{0}$ is along
the horizontal axis.
\end{figure}

\begin{figure}[h]
\epsfxsize=19cm \hspace{-2cm}\epsffile{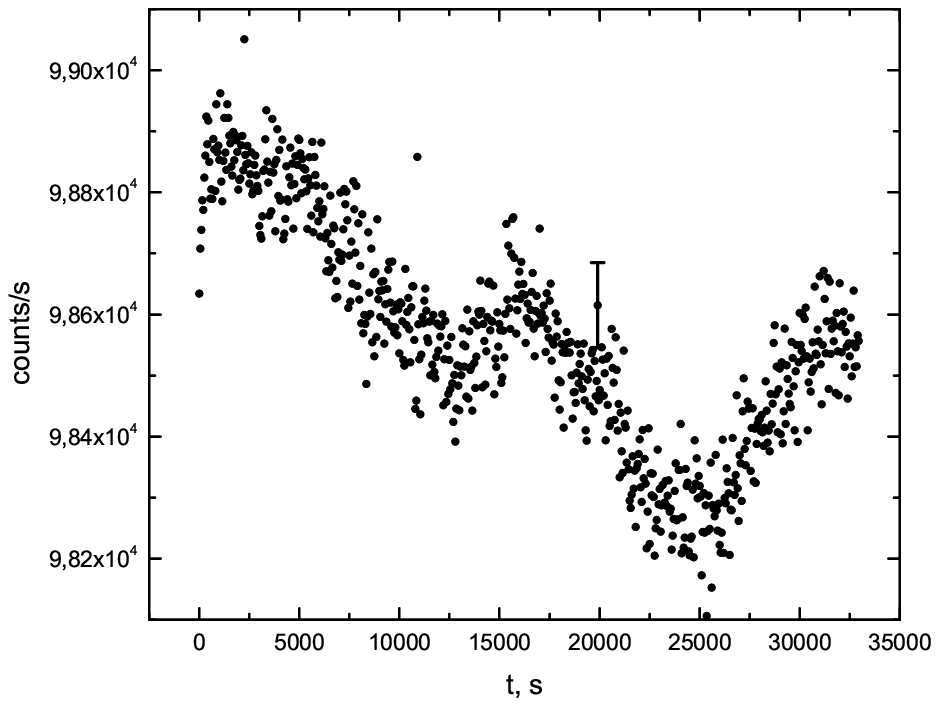}
 Fig. 3. Behavior
of the background on long time scales (revolution 405). Time
relative to the SPI-ACS switch-on after the passage through
radiation belts is along the horizontal axis. A 1$\sigma$
significance is shown for one point.
\end{figure}

\begin{figure}[h]
\epsfxsize=19cm \hspace{-2cm}\epsffile{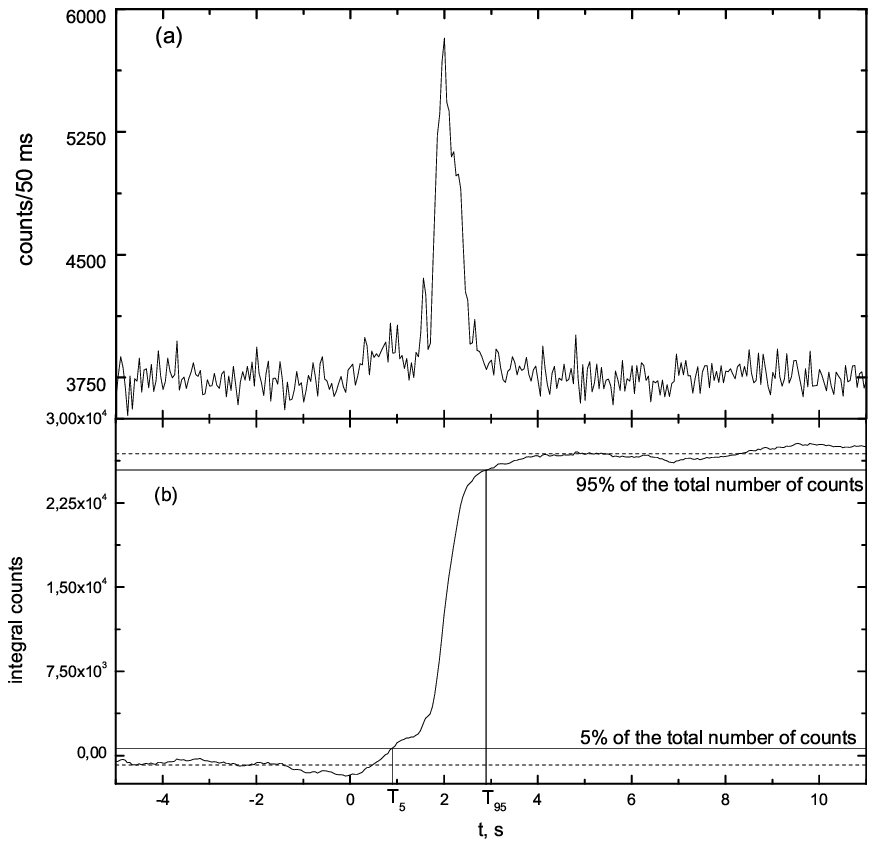} Fig. 4. Scheme
for calculating the duration of GRB 030325: (a) the light curve
with the original time resolution of 50 ms, the number of counts
in 50 ms is along the vertical axis; (b) the corresponding
integral light curve, the integral number of counts after the
background subtraction is along the vertical axis. The horizontal
dotted lines indicate 0 and 100\% levels of the total number of
GRB counts and the solid lines indicate 5 and 95\% levels. The
corresponding values of $T_{5}$ and $T_{95}$ are shown.
\end{figure}

\begin{figure}[h]
\epsfxsize=19cm \hspace{-2cm}\epsffile{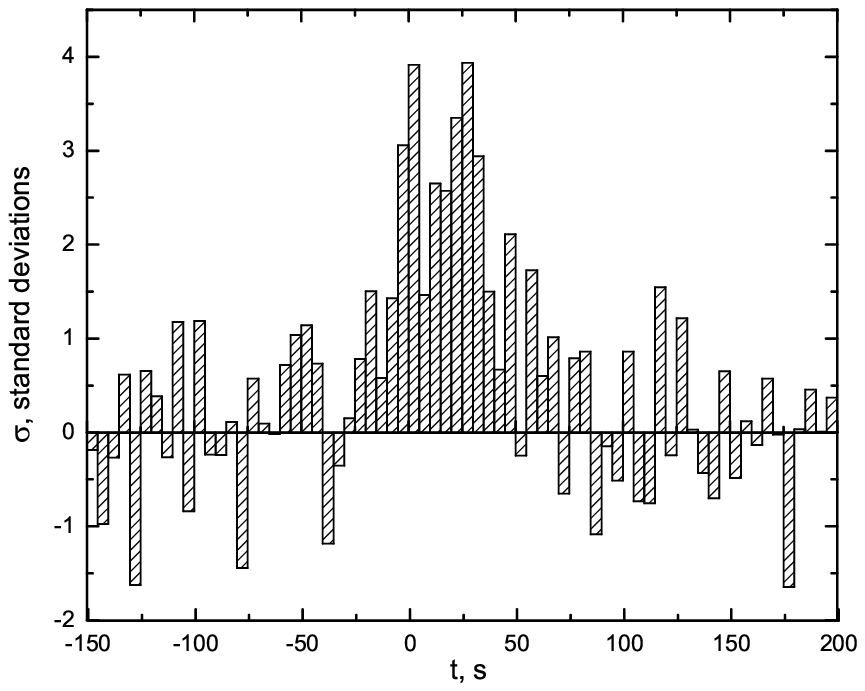} Fig. 5. Result
of processing the light curve for GRB 060221. Time relative to the
trigger $T_{0}$ is along the horizontal axis. The time resolution
is 5 s. Significance in standard deviations is along the vertical
axis. is along the horizontal axis. The time resolution is 5 s.
Significance in standard deviations is along the vertical axis.
\end{figure}

\begin{figure}[h]
\epsfxsize=19cm \hspace{-2cm}\epsffile{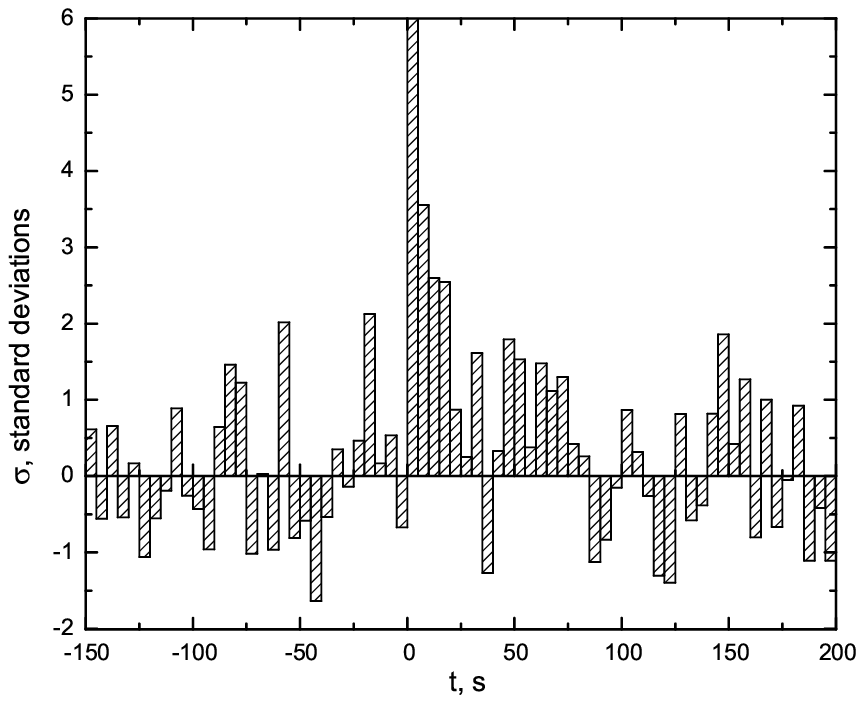} Fig. 6. Same
as
Fig. 5 for GRB 031214. 
\end{figure}

\begin{figure}[h]
\epsfxsize=19cm \hspace{-2cm}\epsffile{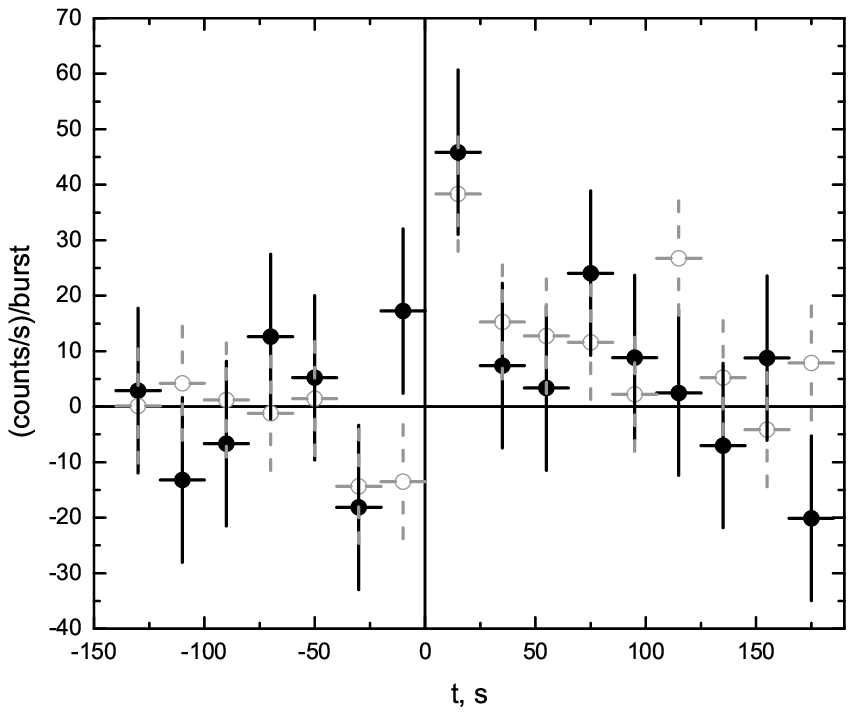} Fig. 7. Averaged
light curve: the filled and open circles represent groups 1 (51
short confirmed GRBs) and 2 (105 unconfirmed short events from the
catalog by Rau et al. 2005), respectively. The 1$\sigma$ errors
are given. The number of counts per second per event is along the
vertical axis. The values corresponding to the primary peak
outside the scale along the vertical axis. 
\end{figure}

\begin{figure}[h]
\epsfxsize=19cm \hspace{-2cm}\epsffile{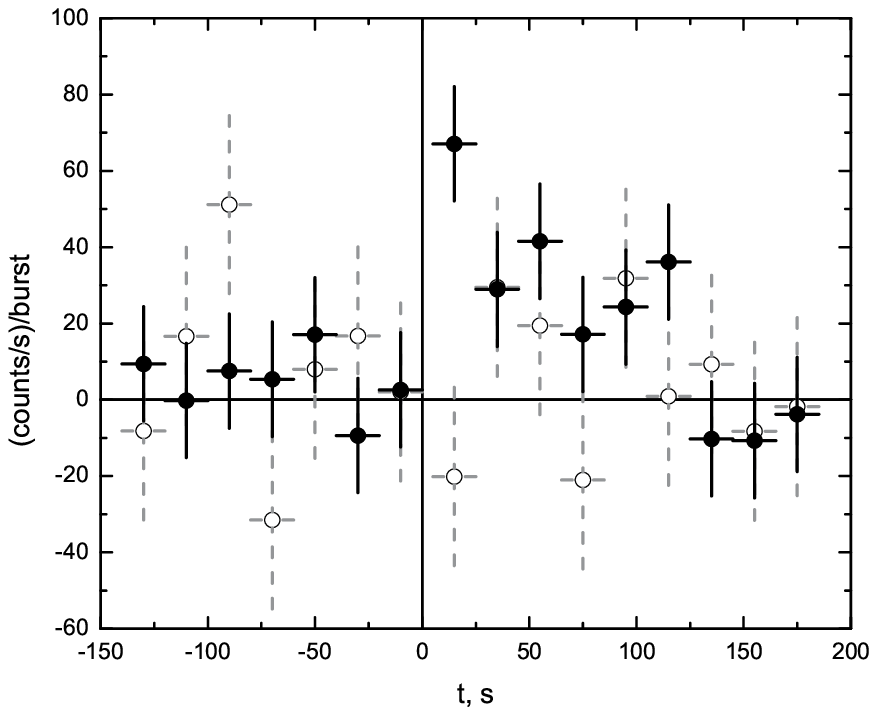} Fig. 8.
Averaged light curve: the filled and open circles represent groups
3 (43 unconfirmed very short events from the catalog by Rau et al.
2005) and 4 (33 unconfirmed events related to triggers from
charged particles), respectively. The 1$\sigma$ errors are given.
The number of counts per second per event is along the vertical
axis. The values corresponding to the primary peak outside the
scale
along the vertical axis.
\end{figure}

\begin{figure}[h]
\epsfxsize=19cm \hspace{-2cm}\epsffile{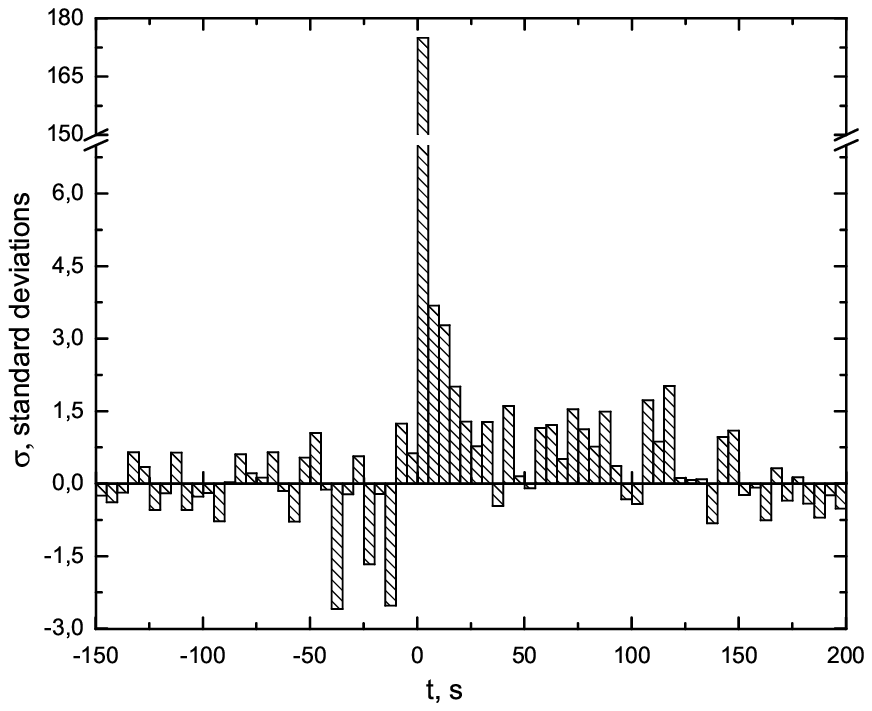} Fig. 9.
Averaged light curve for the combined first and second groups of
events. Time relative to the trigger $T_{0}$ is along the
horizontal axis. The time resolution is 5 s. Significance in
standard
deviations is along the vertical axis.
\end{figure}

\begin{figure}[h]
\epsfxsize=19cm \hspace{-2cm}\epsffile{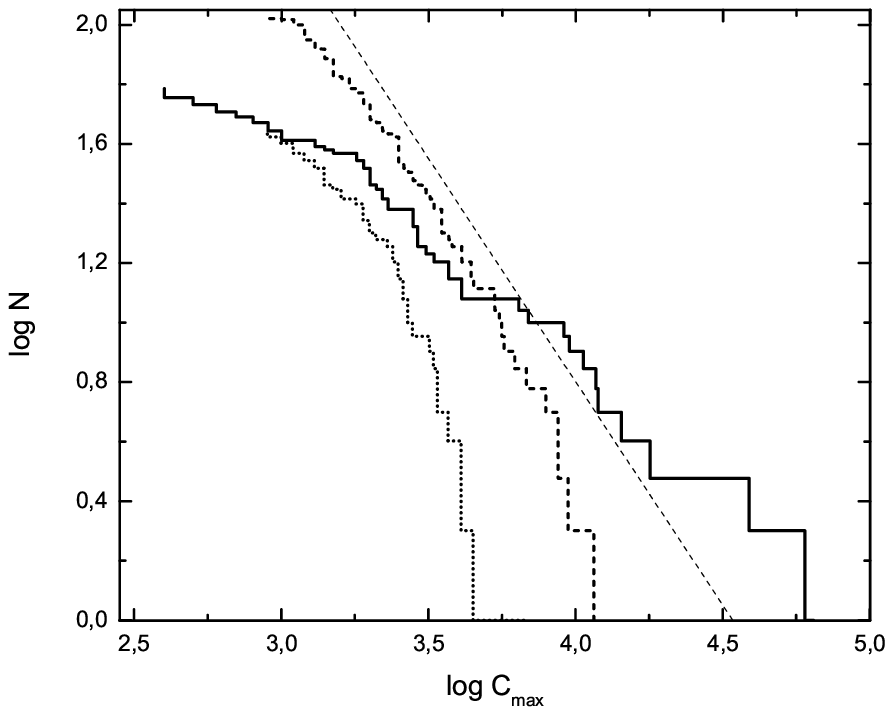} Fig. 10. $log
N-log C_{max}$ distribution: the solid line represents group 1
(confirmed short events from our catalog); the dashed line
represents group 2 (unconfirmed short events from the catalog by
Rau et al. (2005)); the thin dashed line ($log N-log C_{max}\sim$
-3/2) corresponds to a uniform distribution of sources in 3D
Euclidean space.
\end{figure}


\begin{references}

\reference{\it L. Amati}, astro-ph/1002.2232 (2010).

\reference{\it R. A. Burenin}, Astron. Letters {\bf 26}, 323
(2000).

\reference{\it D. B. Cline, C. Matthey, S. Otwinowski}, Nuovo Cim.
{\bf 121B}, 1443 (2006).

\reference{\it V. Connaughton}, \apj\ {\bf 567}, 1028 (2002).

\reference{\it T. Q Donaghy, D. Q. Lamb, T. Sakamoto \etal},
astro-ph/0605570 (2006).

\reference{\it G. Fishman, C. Meegan, R. Wilson \etal}, ApJS {\bf
92}, 229 (1994).

\reference{\it D. D. Frederiks, R. L. Aptekar, S. V. Golenetskii
\etal}, Third Rome Workshop on Gamma-Ray Bursts in the Afterglow
Era (Ed. M. Feroci, F. Frontera, N. Masetti, and L. Piro: San
Francisco: Astronomical Society of the Pacific, 2004) 197.

\reference{\it N. Gehrels, J. P. Norris, V. Mangano \etal}, Nature
{\bf 444}, 1044 (2006).

\reference{\it N. Gehrels}, GAMMA-RAY BURSTS 2007: Proceedings of
the Santa Fe Conference. AIP Conference Proceedings {\bf 1000}, 3
(2008).

\reference {\it W. Hajdas, P. Buhler, C. Eggel \etal}, Astron.
Astrophys. {\bf 411}, L43 (2003).

\reference{\it K. Hurley}
http://www.ssl.berkeley.edu/ipn3/masterli.txt (2008).

\reference{\it A. von Kienlin, V. Beckmann, A. Rau \etal}, Astron.
Astrophys. {\bf 411}, L299 (2003b).

\reference{\it T. M. Koshut, W. S. Paciesas, C. Kouveliotou
\etal}, \apj  {\bf 463}, 570 (1996).

\reference{\it C. Kouveliotou, C. A. Meegan, G. J. Fishman \etal},
\apj\ {\bf 413}, L101 (1993).

\reference {\it A. S. Kozyrev, I. G. Mitrofanov, A. B. Sanin
\etal}, Astron. Lett. {\bf 30}, 435 (2004).

\reference {\it H. Krimm, L. Barbier, S. Barthelmy \etal}, GRB
Coordinates Network {\bf 5704}, 1 (2006).

\reference{\it D. Lazzati, E. Ramirez-Ruiz, G. Ghisellini},
Astron. Astrophys. {\bf 379}, L39 (2001).

\reference {\it Lin Lin, En-Wei Liang, Bin-Bin Zhang \etal},
astro-ph/0809.1796 (2008).

\reference {\it E. P. Mazets, S. V. Golenetskii, V. N. Il'inskii
\etal}, Astrophys. Space Sci. {\bf 80}, 3 (1981).

\reference {\it E. P. Mazets, R. L. Aptekar, T. L. Cline}, \apj\
{\bf 680}, 545 (2008).

\reference {\it S. Mereghetti, D. Gotz, J. Borkowski, R. Walter
and H. Pedersen}, Astron. Astrophys.  {\bf 411}, L291 (2003).

\reference{\it P. Minaev, A. Pozanenko, V. Loznikov}, GAMMA-RAY
BURST: Sixth Huntsville Symposium. AIP Conference Proceedings {\bf
1133}, 418 (2009).

\reference{\it P. Minaev, A. Pozanenko, V. Loznikov},
Astrophysical Bulletin, in press (2010).

\reference{\it I. Mitrofanov, A. Chernenko, A. Pozanenko}, \apj\
{\bf 459}, 570 (1996).

\reference{\it E. Montanari, F. Frontera, C. Guidorzi \etal},
\apj\ {\bf 625}, L17 (2005).

\reference{\it J. P. Norris, J. T. Bonnell}, \apj\ {\bf 643}, 266
(2006).

\reference {\it P. P. Oleynik, S. V. Golenetskii, D. D. Frederiks
\etal}, Physics of Neutron Stars (Ed. D.A. Varshalovich, A.I.
Chugunov, A.Y. Potekhin, and D.G. Yakovlev: Saint-Petersburg:
Saint-Petersburg State Polytechnical University Publishing, 2008)
64.

\reference{\it B. Paczynski}, \apj\ {\bf 494}, L45 (1998).

\reference{\it B. Paczynski}, \apj\ {\bf 308}, L43 (1986).

\reference{\it D. N. Page, S. W. Hawking}, \apj\ {\bf 206}, 1
(1976).

\reference{\it V. Petkov, E. Bugaev, P. Klimai}, Astron. Letters
{\bf 34}, 563 (2008).

\reference{\it A. Rau, A. von Kienlin, K. Hurley \etal}, Astron.
Astrophys. {\bf 438}, 1175 (2005).

\reference{\it J. Ripa, A. Meszaros, C. Wigger \etal}, Astron.
Astrophys. {\bf 498}, 399 (2009).

\reference {\it F. Ryde, L. Borgonovo, S. Larsson \etal}, Astron.
Astrophys. {\bf 411}, L331 (2003).

\reference {\it T. Sakamoto, S. D. Barthelmy, L. Barbier \etal},
\apj\ SS {\bf 175}, 179 (2008).

\reference{\it G. Vianello, D. Gotz, S. Mereghetti},  Astron.
Astrophys. {\bf 495}, 1005 (2009).

\reference{\it D. Vigano, S. Mereghetti}, astro-ph/0912.5329
(2009).


\end{references}
\end{document}